\documentclass[
    ,final            
  ]
  {aipproc}

\layoutstyle{8x11single}

\newcommand\aj{{AJ}}%
\newcommand\apj{{ApJ}}%
\newcommand\apjl{{ApJ}}%
\newcommand\apjs{{ApJS}}%
\newcommand\aap{{A\&A}}%
\newcommand\mnras{{MNRAS}}%
\newcommand\pasj{{PASJ}}%
\newcommand\nat{{Nature}}%
\begin{document}

\title{Type-Ia Supernovae: New Clues to their Progenitors from the
  Delay Time Distribution}

\classification{<Replace this text with PACS numbers; choose from this list:
                \texttt{http://www.aip..org/pacs/index.html}>}

\author{Dan Maoz}{
  address={School of Physics and Astronomy, Tel Aviv University,
  Tel Aviv 69978, Israel}
}

\begin{abstract}
Despite their prominent role in cosmography, little
is yet known about the nature of type-Ia supernovae (SNe Ia), from the identity
of their progenitor systems, through the evolution of those systems
up to ignition and explosion, and to the causes of the 
environmental dependences of their observed properties. I briefly
review some of those puzzles. I then focus on recent progress in
reconstructing the SN Ia delay time distribution (DTD) -- the SN
rate versus time that would follow a hypothetical 
brief burst of star formation. A number of measurements of the DTD
over the past two years, using different methods and based on SNe Ia
in different environments and redshift ranges, are converging.
At delays $1<t<10$~Gyr, these measurements show
a similar $\sim t^{-1}$ power-law shape, with similar normalizations.
The DTD peaks at the shortest delays probed, but there is still some
uncertainty regarding its precise shape in the range $0.1<t<1$~Gyr. 
At face value, this result supports
Ron Webbinks's (1984) idea of a double-degenerate 
progenitor origin for SNe~Ia, but the
numbers currently predicted by binary population synthesis models must be
increased by factors of 3-10, at least. Single-degenerate progenitors
may still play a role in producing short-delay SNe Ia, or perhaps
all SNe Ia, if there are fundamental errors in the current
modeling attempts.
\end{abstract}

\maketitle


\section{What we do not understand about type-Ia supernovae}
\label{ss.intro}
Supernovae (SNe) play a central role in astrophysics, not only as
distance indicators for cosmology (e.g., Riess et al. 1998; Perlmutter
et al. 1999), but as prime synthesizers of heavy
elements (e.g. Woosley 2007), 
sources of kinetic energy, and accelerators of cosmic rays
(e.g. Helder et al. 2009). 
However, many of the most basic facts about these events are still
poorly understood.  
Type-Ia SNe (SNe~Ia), the subject of this review, are linked by indirect
evidence to the thermonuclear detonations of  carbon-oxygen white
dwarfs (WDs) whose masses have grown to near the Chandrasekhar limit
(Hoyle \& Fowler 1960). However,
competing scenarios exist for the initial conditions and evolutionary 
paths that lead to this mass growth. In the single degenerate (SD)
model (Whelan \& Iben 1974), a WD grows in mass through accretion from
a non-degenerate stellar companion. The 
double degenerate (DD) scenario, on the other hand, was first proposed
in the landmark paper by Ron (Webbink 1984), the same 
paper we have seen referenced in so many of the talks at this meeting,
on so many different topics. (The same idea was proposed simultaneously by  
Iben \& Tutukov 1984). In this scenario, 
two WDs merge after losing energy and 
angular momentum to gravitational waves. To cite Ron in the abstract of
that paper:
``C/O-C/O pairs are again unstable to dynamical time-scale mass
transfer and, since their masses exceed the Chandrasekhar limit, are
destined to become SNe''.

Although 27 years have passed, 
 neither the SD nor the DD models can yet be excluded observationally (nor clearly
 favored), and hence we find ourselves in an embarassing situation: 
these explosions are of the utmost astrophysical importance, we use 
them to reach radical conclusions about the existence and properties
of dark energy, and yet we do not even know for sure {\it what} is exploding!
Indeed, the recent Astronomy and Astrophysics 
Decadal Survey
 (\url{http://sites.nationalacademies.org/bpa/BPA_049810}) 
has listed 
determining the nature of SN Ia progenitors as one of the major
objectives of the coming decade.
Apart from the wide-open progenitor question,
 many questions remain  
regarding the subsequent phases: the details of the probably multiple
accretion episodes; the enigmatic common-envelope phase; the
type and the location of the ignition; the nature of the combustion -- 
deflagration, detonation, or both; and within all of these details,
the eventual causes of the regularities, trends, and environmental
dependences observed in SN Ia light curves and spectra (see, e.g., the
recent review by Howell 2010). Interestingly, even the
near-Chandrasekhar-mass conjecture has come under renewed scrutiny.
Sub-Chandrasekhar explosions have been proposed as a way of explaining
some, or perhaps even most SN Ia events (Raskin et al. 2009; Rosswog et
al. 2009; Sim et al. 2010; van Kekwijk, Chang \& Justham
2010). Conversely, the Ni mass deduced for some SN Ia explosions is
strongly suggestive of a super-Chandrasekhar-mass progenitor
 (e.g. Tanaka et al. 2010; Silverman et al. 2010).

Both the SD and the DD models suffer from numerous problems, both
theoretical and observational. In terms of SD theory, it has long been 
recognized that the mass accretion rate on to the WD needs to be
within a narrow range, in order to attain stable hydrogen burning on the
surface, and mass growth toward the Chandrasekhar mass. 
Too-low an accretion rate will lead to explosive
ignition of the accreted hydrogen layer in a nova event, which likely
blows away more material from the WD than was gained (e.g. Townsley
\& Bildsten 2005). Too-high an
accretion rate will lead to the escape of the accreted material in a
wind. The self-regulation of the accretion flow by the wind from the
companion, as conceived by Hachisu \& Nomoto (e.g. Hachisu et al. 1999) has
thus long been considered to be an essential element of the SD model. 
Questions, however, have been raised as to whether 
the mechanism does not 
require too much fine tuning (Cassisi et al. 1998; Piersanti et
al. 2000; Shen \& Bildsten 2007).

The SD model faces additional obstacles when it comes to observational
searches for its signatures. Badenes et al. (2007) searched seven
young SN Ia remnants for the wind-blown cavities that would be
expected in the wind-regulation picture. Instead, in every case it
appeared the remnant is expanding into a constant-density ISM. Leonard
(2007) obtained deep spectroscopy in the late nebular phase of several SNe
Ia, in search of the trace amounts of H or He that would be expected
from the stellar winds. None was found. Prieto et al. (2008) have
pointed out the SNe Ia have been observed in galaxies with quite low 
metallicities. This may run counter to the expectations that, at low
enough metallicities, the optical depth of the wind would become
small, and the hence the wind-regulation mechanism would become
ineffective. Variable NaD absorption has been detected in the spectra
of a few SNe Ia (Patat et al. 2007; Simon et al. 2009) and has been
interpreted as circumstellar material from the companion in an SD
model. But why is such absorption seen in only a minority of cases searched?
The companion, in an SD scenario, will survive the explosion, and is
likely to be identifiable by virtue of its anomalous velocity,
rotation, spectrum, or composition. However, searches for the survivor 
of Tycho's SN have not been able to reach a consensus
(Ruiz Lapuente et al. 2004; Fuhrman 2005;
Ihara et al. 2007; Gonzalez-Hernandez et al. 2009; Kerzendorf et
al. 2009). Perhaps the effects of the explosion on the companion are
more benign than once thought (see Pakmor et al. 2008). Hayden et
al. (2010), however, place observational limits on the presence
 of any shock signatures in the light curves of 108 SNe Ia with good
 early-time coverage, shocks that are expected from the ejecta hitting the
 companion, as calculated by Kasen (2010).
 Finally, di Stefano (2010), and Gilfanov \&
Bogdan (2010), have raised related arguments that the accreting and
growing WDs in the SD scenario would be undergoing stable nuclear
burning on their surfaces, and hence would be visible as super-soft
X-ray sources (SSS). The actual numbers of SSS are far below those 
required to explain the observed SN Ia rate.

The DD model is also not free of problems. Foremost, it has long
been argued  that the merger of two unequal-mass WDs will lead to 
an accretion-induced collapse and the formation of a neutron star,
i.e. a core-collapse SN, rather than a SN Ia (Nomoto \& Iben 1985;
Guerrero et al. 2004). Others, however, have argued for ways to avoid
this outcome (Piersanti et al. 2003; Pakmor et al. 2010; 
Van Kerkwijk et al. 2010).
Observationally, it has been much harder to find evidence either for
or against the DD scenario because, by construction, it leaves
essentially no traces -- it is the perfect crime! The most promising 
avenue has been to search the Solar neighborhood for the close and
massive WD binaries that will merge within a Hubble time, surpassing
the Chandrasekhar mass and presumably 
producing DD SNe Ia. The largest survey to
date, SPY (Napiwotzki et al. 2004; Geier et al. 2007) has found no
such pairs among $\sim 1000$ WDs. However, only of order one pair is probably 
expected given the (highly uncertain) Galactic SN Ia
rate (Maoz 2008). Clearer results may emerge from the ongoing SWARMS survey by 
Badenes et al. (2009), which searches for close binaries among an
order of magnitude more WDs in the Sloan Digital Sky Survey.

There are additional problems that are shared by both scenarios, SD
and DD. The energetics and spectra of the explosions do not come out
right, unless finely (and artificially) tuned in an initial subsonic
deflagration that, at the right point in time, spontaneously evolves into
a supersonic detonation (Khokhlov 1991). 
If the ignited mass is always near-Chandrasekhar, why is there the 
range of SN Ia luminosities inherent to the Phillips (1993) relation?
(Indeed, Sim et al. 2010 show that their sub-Chandrasekhar models do a fair
job of reproducing the Phillips relation).
Why is there a dependence of the SN Ia luminosity (or, equivalently, 
the mass of radioactive Ni synthesized) on the age of the galaxy host
(e.g. Howell et al. 2009)?
The oldest hosts, with little star formation, are clearly observed to
host only faint, low-stretch, SNe Ia, while star-forming galaxies host
both bright-and-slow and faint-and-fast SNe Ia. Finally, both scenarios
predict, based on binary population synthesis, SN rates that are lower
than actually observed (more on this later). 

\section{The delay-time distribution}
A fundamental function that can shed light on the progenitor question is    
the SN delay time distribution (DTD). The DTD is the hypothetical SN
rate versus time that would follow a brief burst of star formation. 
The DTD is directly linked to the lifetimes (i.e.,
the initial masses) of the progenitors and to the 
binary evolution timescales up to the
explosion, and therefore different progenitor scenarios predict
different DTDs. 
Various theoretical 
forms have been proposed for the DTD, some derived from detailed
binary population synthesis calculations
(e.g., Yungelson \& Livio 2000;  Han \& Podsiadlowski
2004; Ruiter et al. 2009;
Mennekens et al. 2010; see also contributions to this volume by
Claeys, Mennekens,
Ruiter, Toonen, and Wang);
some physically motivated mathematical parameterizations, with varying
degrees of sophistication (e.g., Madau
et al. 1998; Greggio 2005; Totani et al. 2008); 
and some ad hoc formulations intended to
reproduce the observed field SN rate evolution (e.g., Strolger et
al. 2004).

 Until recently, only few, and
often-contradictory, observational constraints on the DTD existed. In
the
past year or so, the observational situation is changing dramatically,
and a clear view of the DTD is emerging, one that is beginning to
discriminate among SN Ia progenitor models.
I review these observations, with emphasis on the most recent ones.
   
\subsection{SN Ia rates versus redshift, compared to cosmic
  star-formation history}
One observational 
approach to recovering the DTD 
has been to compare the SN rate in field galaxies, as a
function of redshift, to the cosmic star formation history (SFH). Given that
the DTD is the SN ``response'' to a short burst of star formation, 
the SN rate versus cosmic time, $R_{Ia}(t)$, 
will be the convolution of the DTD with the SFH (i.e. the 
star formation rate versus cosmic time), $S(t)$,
\begin{equation}
R_{Ia}(t)\propto\int_{0}^{t}S(t-\tau)\frac{\Psi(\tau)}{m(\tau)}d\tau
,
\end{equation}
where $m(\tau)$ is the surviving mass fraction in a stellar
population, after accounting for the mass losses during stellar
evolution due to SNe and winds (and is obtainable from stellar population
synthesis models).
Here and throughout, we will be
considering SN rates measured per unit stellar mass {\it at the time of
observation}, and DTDs normalized per unit stellar mass {\it
  formed}. In making intercomparisons of measurements among
themselves, and with predictions, I have taken special care that consistent
definitions and  stellar initial mass functions (IMFs) be assumed.

Gal-Yam \& Maoz (2004) carried
out the first such comparison, using a small sample of SNe~Ia out to
$z=0.8$, and concluded that the results were strongly dependent on the
poorly known cosmic SFH, 
a conclusion echoed by Forster et al. (2006).
 With the availability of SN rate measurements
to higher redshifts, Barris \& Tonry (2006) found a SN~Ia rate that
closely tracks the SFH out to $z\sim 1$, and concluded  that the DTD
must be concentrated at short delays, $<1$~Gyr. Similar
conclusions have been reached, at least out to $z\sim 0.7$, by
Sullivan et al. (2006). In contrast,
Dahlen et al. (2004, 2008) and Strolger et al. (2004, 2010) 
have argued for a DTD
that is peaked at a delay of $\sim 3$~Gyr, with little power at short
delays, based on a sharp decrease in the SN~Ia rate at $z>1$ found by
them in the HST/GOODS survey. However,
Kuznetzova et al. (2007) re-analyzed some of these datasets and
concluded that the small numbers of SNe and their potential
classification  errors preclude reaching a conclusion. Similarly,
Poznanski et al. (2007) performed new measurements of the $z>1$ SN~Ia
rate by surveying the Subaru Deeep Field with the Subaru
SuprimeCam. We
 found that, within uncertainties, the SN rate could be
tracking the SFH. This, again, would imply a short delay time.
Greggio et al. (2008) pointed out that underestimated extinction
of the highest-$z$ SNe, observed in their rest-frame ultraviolet
emission, could be an additional factor affecting these results.
Blanc \& Greggio (2008) and Horiuchi \& Beacom (2010)  have shown that, within the errors, a wide
range of DTDs is consistent with the data, but with a preference for
a DTD similar to $\sim t^{-1}$. 
\begin{figure}
  \includegraphics[height=.3\textheight]{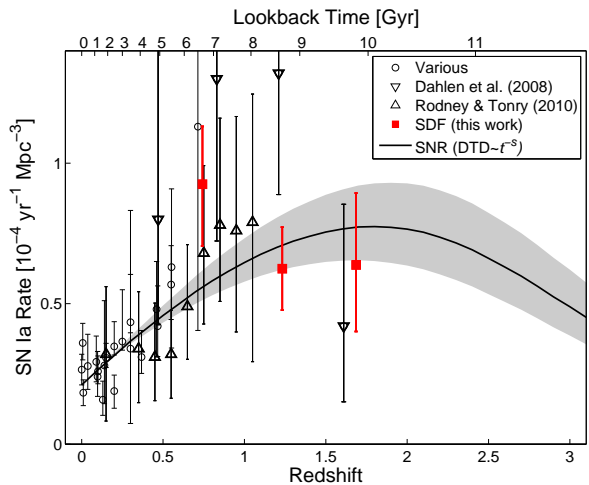}
  \caption{SN Ia rate versus redshift. Filled squares (red) 
are from the Subaru
  Deep Field search by Graur et al. (in preparation). The curve is
  obtained by convolving the SFH of Hopkins \& Beacom (2006) with a
  DTD of the form $\Psi(t)\propto t^{-1}$.
}
\end{figure}

Happily, at the time of this writing, it appears that the picture
is finally clarifying and converging with respect to the field SN Ia rate
as a function of redshift, and the DTD that it implies. 
Rodney \& Tonry (2010) have presented a 
re-analysis of the data of Barris \& Tonry (2006), with new SN Ia
rates that are lowered, and in much better agreement with other measurements 
at similar redshifts. Preliminary new rates from the Supernova Legacy 
Survey (Perrett et al., in preparation) agree with the revised numbers, and suggest a SN Ia rate
that continues to rise out to $z=1$, albeit growing more gradually
than the SFH. Finally, a quadrupling of the initial high-$z$ SN sample,
first presented by Poznanski et al. (2007), is resolving the puzzle of the
SN rate out to $z=2$. Graur et al. (in preparation) present a sample
of 150 SNe discovered by ``staring'' at a single Subaru SuprimeCam
field -- the Subaru Deep Field -- at four independent  epochs, 
with 2 full nights of integration per epoch. SN host galaxy redshifts
are based on spectral and photometric redshifts, from the
extensive
UV to IR database existing for this field. Classification of the SN
candidates is photometric. The SN sample includes 26 events that are 
fully consistent with being normal SNe Ia in the redshift range
$1.0<z<1.5$, and 10-12 such events at $1.5<z<2.0$. 
The rates derived from the Subaru data merge
 smoothly with the most recent and 
most accurate rate measurements at $z<1$, confirming the trend of
a SN Ia rate that gradually levels off at high $z$, but does not dive
down, as previously claimed by Dahlen et al. (2004, 2008).
In Graur et al., we find that a DTD with a
power-law form, $\Psi(t)\propto t^{-1}$, when
convolved with the Hopkins \& Beacom (2006) SFH, gives an excellent 
fit to the observed SN rates. The rates and this fit are shown 
in Fig.~1.

\subsection{SN Ia rate versus galaxy ``age''}
Another approach to recovering the DTD has been to compare the SN
rates in galaxy populations of different characteristic ages. Using
this approach, Mannucci
et al. (2005, 2006), Scannapieco \& Bildsten (2005), and Sullivan
(2006) all found evidence for the co-existence of two SN~Ia
populations, a ``prompt'' population that explodes within  $\sim
100-500$~Myr, and a delayed channel that produces SNe~Ia on timescales of
order 5~Gyr. This has led to the ``$A+B$'' formulation which, in
essence, is just a DTD with two coarse time bins. The $B$ parameter,
divided by the assumed duration of the prompt component, is the mean
SN rate in the first, prompt, time bin of the DTD. The $A$ parameter
is (after correcting for stellar mass loss, $m(t)$)
the
mean rate in the second, delayed, time bin.
Naturally, these two ``channels'' may in reality be just 
integrals over a continuous DTD on two sides of some time border
(Greggio et al. 2008). 
Totani et al. (2008) have used a similar approach to recover the DTD, by
comparing SN~Ia rates in early-type galaxies of different
characteristic ages, seen at $z=0.4-1.2$ as
part of the Subaru/XMM-Newton Deep Survey (SXDS) project.
They find a DTD consistent with a $t^{-1}$ form. 
 Additional 
recent attempts to address the issue with the ``rate vs. age''
approach have been made by Aubourg et al. (2008), Raskin et al. 
(2009), Yasuda \& Fukugita
(2009), and Cooper et al. (2009).

\subsection{SN Ia rate versus individual galaxy star formation histories}
Both of the approaches described above involve averaging, and hence some loss
of information. In the first approach, one averages over large galaxy
populations, by associating all of the SNe detected at a given
redshift with all the galaxies of a particular type at that redshift.
\begin{figure}
  \includegraphics[height=.3\textheight]{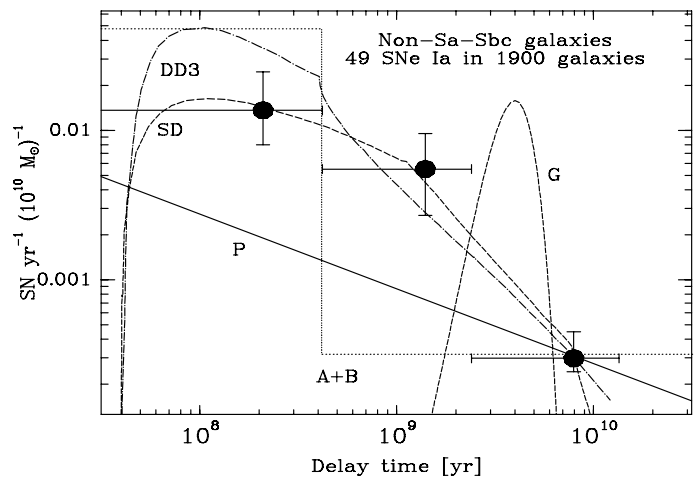}
  \caption{DTD recovered by Maoz et al. (2010a) from a subsample of
  the Lick Observatory SN Search galaxies and their SNe, based on the
individual SFH of each galaxy from its SDSS spectrum. The example
  shown uses a subsample excluding intermediate Hubble types, which
  are prone to incorrect representation of the full SFH of the galaxy,
due to the limited aprture of the SDSS spectrograph fibers. The
  observed DTD is compared to several theoretical and observed 
DTDs: the $t^{-0.5}$
  power law proposed by Pritchet et al. (2008; P); the Gaussian DTD found
by Dahlen et al. (2004, 2008; G); SD and DD analytic models by Greggio
  (2005); and the DTD implied by the A+B picture of Scannapieco \&
  Bildsten (2005).
 }
\end{figure}
In the second approach, a characteristic age for a sample of
galaxies replaces the detailed SFH of the individual galaxies in a SN survey.
Maoz et al. (2010a) recently presented 
a method for recovering the DTD, which avoids this averaging.
In the method, the SFH of every individual galaxy, or even galaxy
subunit, is compared to the number of SNe it hosted in the survey
(generally none, sometimes one, rarely more). DTD recovery is treated
as a linearized inverse problem, which is solved statistically.
Maoz et al. (2010a) applied the method to a subsample of the 
galaxies, and the SNe they hosted, in the 
Lick Observatory SN Search (LOSS; Leaman et
al. 2010; Li et al. 2010a,b). This has been that largest survey for
local (<200~Mpc) SNe over the past 15 years. From the 15,000 LOSS
survey galaxies, we chose subsamples 
 having spectral-synthesis-based SFH reconstructions  by Tojeiro et
al. (2009), based on spectra from the Sloan
Digital Sky Survey (SDSS; York et al, 2000). In the recovered DTD
(Fig.~2) Maoz et al. (2010a)
 find a significant detection of both
a prompt SN~Ia component, that explodes within 420~Myr of star formation, 
and a delayed SN~Ia with population that explodes after $>2.4$~Gyr. 
A related DTD reconstruction method has been applied by Brandt et
al. (2010) to the SNe~Ia from the SDSS II survey. Like Maoz et
al. (2010a), they detect both a prompt and a delayed SN~Ia population.

\subsection{SN remnants in nearby galaxies with SFHs based on resolved
  stellar populations}
Another application of the idea to reconstruct the DTD while taking
 into account SFHs, rather than mean ages, was made
by Maoz \& Badenes (2010). We 
applied this method to a sample of SN remnants in the Magellanic
Clouds, which we compiled in Badenes, Maoz, \& Draine (2010). 
The Clouds have very detailed SFHs in many small individual spatial
cells, obtained by Harris \& Zaritsky (2004, 2009), by fitting model stellar
isochrones to the resolved stellar populations. Thus, one can compare
the SFH of each individual cell to the number of SNe it hosted (or did not)
over the past few kyr, as evidenced by the observed remnants. This
turns the remnants in the Clouds into
an effective SN survey, although several complications need to be
dealt with (see Badenes et al. 2010 and Maoz \& Badenes 2010).  
Using this method, we again find 
a significant detection of a prompt (this time $<330$~Myr) SN~Ia
component. We are currently producing larger samples by using the
SN remnant
populations in additional nearby galaxies, such as M33 and M31, and
their spatially resolved SFHs, again based on the resolved stellar populations.

A frequent objection that arises, when considering this approach, is
that one cannot correctly deduce SN delay times by comparing, on the
one hand, star
formation rates in a small projected piece of a galaxy to, on the
other hand, the SNe that
this region of the galaxy is seen to host, since random velocities cause the SN
progenitor, by the time it explodes, to have drifted far from its
birth location. While this objection is indeed valid if one is comparing 
SNe to the mean ages of their locations, it does not apply if, as
here, we are considering detailed SFHs (rather than mean ages), for
full ensembles of galaxy cells and SNe. The reason is that both the
SN progenitors and their entire parent populations undergo the same
spatial diffusion within a galaxy over time. This is explained in more
detail in Maoz et al. (2010a) and Maoz \& Badenes (2010).
 
\subsection{SN Ia rates versus redhsift in galaxy clusters}
The last approach I will review for recovering the DTD
  is to measure the SN rate
vs. redshift in massive galaxy clusters.
The deep potential wells of clusters, combined with their
relatively simple SFHs, make them ideal locations for
studying both the DTD and the metal production of SNe. 
Optical spectroscopy and multiwavelength photometry 
of cluster galaxies has shown consistently that the bulk of their
stars were formed within short episodes ($\sim 100$~Myr) 
at $z\sim 3$ (e.g., Daddi et al. 2000; Stanford et al. 2005;
 Eisenhardt et al. 2008). Thus, the
observed SN rate vs. cosmic time $t$, given  
a stellar formation epoch $t_f$,  
provides an almost direct
measurement of the form of the DTD,
\begin{equation}
R_{Ia}(t)=\frac{\Psi(t-t_f)}{m(t-t_f)}.
\label{ddtmassloss}
\end{equation}
Furthermore, the record of metals stored in   
the intracluster medium (ICM) constrains the  integrated 
number of SNe Ia per formed stellar mass, $N_{\rm SN}/M_*$, that have
exploded in the cluster over its stellar age, $t_0$, 
and hence the normalization of the DTD,
\begin{equation}
\int_0^{t_0} \Psi (t) dt=\frac{N_{\rm SN}}{M_*}.
\label{normconstraint}
\end{equation} 
As reviewed in detail in Maoz et al. (2010b), X-ray and optical
observations of galaxy clusters have reached the point where they 
constrain $N_{\rm SN}/M_*$ to the level of $\pm 50\%$, based
on the observed abundances of iron (the main product of SN Ia
explosions), after accounting for the contributions by core-collapse
SNe (and the uncertainty in that contribution).
 \begin{figure}
  \includegraphics[height=.3\textheight]{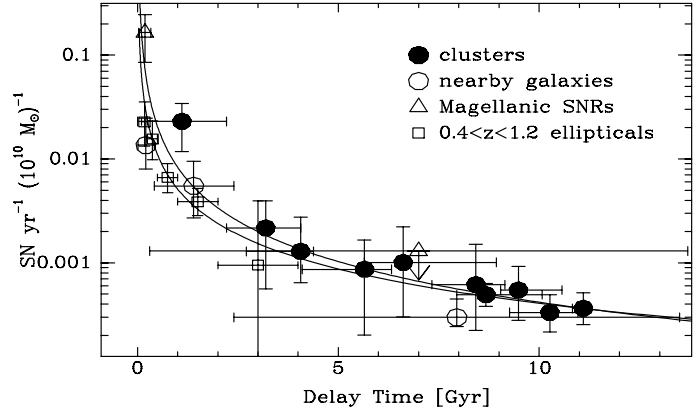}
  \caption{Filled points: 
SN Ia DTD recovered based on galaxy cluster SN Ia rate measurements,
and cluster iron abundances, from Maoz et al. (2010b). Also shown
are some of the DTD derivations previously described, using other
methods, in different environments and different redshifts. The solid
curves are power laws, $t^{-1.1}$ and $t^{-1.3}$,
 that describe well these results, as well as
the latest field SN Ia rate measurements out to $z=2$, when compared
to the cosmic SFH (see Fig.~1).
}
\label{clusterratefig}
\end{figure}
\begin{figure}
  \includegraphics[height=.3\textheight]{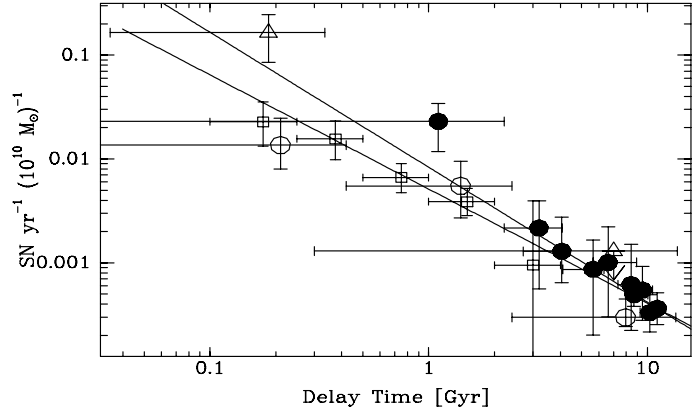}
  \caption{Same as Fig.~\ref{clusterratefig}, but with a logarithmic
  time axis}.
\label{clusterratefiglog}
\end{figure}

A decade ago, there were no real measurements of SN rates in galaxy clusters. 
However, 
the observational situation has again improved dramatically,
especially in the last few years. Following large investments of
effort
and observational resources,
fairly accurate  cluster SN~Ia rates have now been measured 
in the redshift range
from $0<z<1.5$ (Gal-Yam et al. 2002,2008; Sharon et al. 2007, 2010;
Mannucci et al. 2008; Graham et al. 2008; Dilday et al. 2010; Barbary
et al. 2010). 
Figure~\ref{clusterratefig} shows (filled points) the DTD derived by Maoz
et al. (2010b) based on these galaxy-cluster SN Ia rate
measurements, together with the iron-based DTD integral constraint,
which sets the level in the earliest DTD bin. Also plotted, with open
symbols,
 are some
of the recent DTD measurements described previously: the DTD from the ages
of high-$z$ field ellipticals (Totani et al. 2008); the DTD from the nearby
LOSS galaxies and their SDSS-based SFHs (Maoz et al. 2010a, Fig.~2); the DTD
from the Magellanic Cloud SN remnants by Maoz \& Badenes (2010); and 
(solid curves) the power-law DTDs that best fit the high-$z$ field SN
rates in the Subaru Deep Field, by Graur et al., when compared to the
cosmic SFH (see Fig.~1). Figure~\ref{clusterratefiglog} shows the same data, but 
on a logarithmic time axis that illustrates more clearly the situation at
short time delays.

 \section{Synthesis}
The picture emerging from
Figs.~\ref{clusterratefig}-\ref{clusterratefiglog}
is remarkable. For one, all of these diverse DTD determinations, based
on different methods, using SNe Ia in different environments and
at different redshifts, agree with each other, both in form and in absolute
level. At delays $t>1$~ Gyr, there seems little doubt that the DTD is 
well described by a power law of the form $t^{-s}$, with $s\approx 1$.
The union of the 95\% confidence ranges resulting from
the detailed statistical analyses in the papers above is $0.9<s<1.5$. 
At delays $t < 1$~Gyr, the picture is not as clear cut. Nonetheless, it
{\it is} clear that the DTD does peak in that earliest time bin.
It may continue to rise to short delays with the same slope seen at 
long delays, or it may transit to a shallower rise, but it certainly
does not fall. The explosion of at least $\sim 1/2$ of SNe Ia 
within 1~Gyr of star formation is, by now, probably an inescapable fact.

Apart from the form of the DTD, there is also fairly good agreement, among all
the derivations, on its normalization, or equivalently, its integral
over a Hubble time. Once due attention is given to consistent
definitions and assumptions of IMF (see above), 
the time-integrated number of SNe Ia per stellar mass formed is in the
range of $N_{\rm SN}/M_*=(1.5-3.5)\times 10^{-3}~{\rm M_\odot}^{-1}$,
assuming a realistic IMF, with a turnover at low stellar masses.
  
What does this observed DTD imply for the burning questions on the progenitors
of SNe Ia? 
Power laws
 have been long considered as possible
forms of the DTD (e.g., Ciotti et al. 1991; Sadat et
 al. 1998). As noted by previous authors (e.g., Greggio 2005;
Totani et al. 2008) a power-law dependence
is generic to models (such as the DD model) in which
the event rate ultimately depends on the loss of energy and angular
momentum
to gravitational radiation by the progenitor binary system.
If the dynamics are controlled solely by gravitational wave losses,
the time $t$ until a merger depends on the
binary separation $a$ as
\begin{equation}
t\sim a^4.
\end{equation}
If the separations are distributed as a power law
\begin{equation}
\frac{dN}{da}\sim a^\epsilon,
\end{equation}
then the event rate will be
\begin{equation}
\frac{dN}{dt}=\frac{dN}{da}\frac{da}{dt}\sim t^{(\epsilon -3)/4} .
\label{DDdependence}
\end{equation}
For  a fairly large range around $\epsilon\approx -1$, which describes
well the observed distribution of initial separations  of
non-interacting
binaries (see Maoz 2008 for a review of the issue in the present
context),
the DTD will
have a power-law dependence with index not far from $-1$.
Indeed, a $\sim t^{-1}$ power law appears to be a generic outcome also of 
detailed binary population synthesis calculations of the DD channel
(e.g., Yungelson \& Livio 2000;   
Mennekens et al. 2010; see Toonen, Claeys in this volume).
However, in
reality, the binary separation distribution of WDs that have
emerged from their common envelope phase could be radically different,
given the complexity of the physics of that phase. Thus, the $\sim            
t^{-1}$ DTD dependence of the DD channel cannot be considered
unavoidable.
Be that as it may, the observed DTD reconstructions 
 all point to a $\sim t^{-1}$ power-law.

A different power-law DTD dependence, with different physical
 motivation, has
been proposed by Pritchet et al. (2008). If the time between formation
of a WD and its explosion as a SN~Ia is always brief compared to the
formation time of the WD, the DTD will simply be proportional to the
formation rate of WDs. Assuming that the main-sequence lifetime of a
 star
depends on its initial mass, $m$, as  a power law,
\begin{equation}
t\sim m^\delta,
\end{equation}
and assuming the IMF is also a power law,
\begin{equation}
\frac{dN}{dm}\sim m^\lambda,
\end{equation}
then the WD formation rate, and hence the DTD, will be
\begin{equation}
\frac{dN}{dt}=\frac{dN}{dm}\frac{dm}{dt}\sim
t^{(1+\lambda-\delta)/\delta} .
\end{equation}
 For the commonly used value of $\delta=-2.5$ and the Salpeter
 (1955) slope of $\lambda=-2.35$, the resulting power-law index is
 $-0.46$, or roughly $-1/2$. Pritchet et al. (2008) raised the 
possibility
of such a $t^{-1/2}$ DTD.  
It is arguable that, instead of
 a single, $\sim t^{-1}$ power law, motivated by binary mergers,
with this power law
  extending back to delays as short as
 40~Myr (the lifetime of the most massive stars that form WDs), 
there could be a ``bottleneck'' in the supply of progenitor
systems below some delay. Such a bottleneck could be
 due to the birth rate of WDs, which
behaves as $\sim t^{-1/2}$. One possible result would then be a
 broken-power-law DTD, with $\Psi\propto t^{-1/2}$ up to some
  time, $t_c$, and $\Psi\propto t^{-1}$ thereafter.
A possible value could be  $t_c\approx 400$~Myr, corresponding to the
lifetimes of $3M_\odot$ stars. If that were the lowest initial mass of
 stars that can produce the WD primary in a DD SN~Ia progenitor, then
beyond $t_c$ the supply of new systems would go to zero, and the
SN~Ia rate would be dictated by the merger rate. Indeed, the Greggio
 (2005) DD model, shown in Fig.~2, above, is essentially a
 $t^{-1/2}$,~$t^{-1.3}$, broken power-law with break at
$t_c<400$~Myr. The detailed fits to the various observational DTDs
show that such broken power laws are acceptable, 
as long as $t_c<1.5$~Gyr.
Such a late break time is interesting
in the context of sub-Chandra merger models, in which the mergers
of white dwarfs that had main sequence masses smaller than $3{\rm M_\odot}$
can produce SNe~Ia (Sim et al. 2010; Van Kerkwijk et al. 2010).

In terms of the observed DTD normalizations, the DD models do not fare
as well. As already noted by Maoz (2008), Ruiter et
al. (2008), Mennekens et al. (2010), and Maoz et al. (2010b), 
binary synthesis DD models underpredict
observed SN rates by factors of at least a few, and likely by
more. It is not clear at present if there is a good way to alleviate this
inconsistency with the observations.
 
While the DTD predicted by Ron's DD model appears to be supported by
the observations, at least in terms of form, if not normalization, the
situation is quite different for the SD scenario. There is a
staggering variety of results among the predictions for the DTD from 
SD models. Some of this variety is due to the fact that ``SD''
includes an assortment of very different sub-channels. Some of it is
due to the fact that, even within a given sub-channel, different
workers treat the same evolutionary phases using different
approximations
 (e.g. the 
common-phase phase, via Ron's $\alpha$ formalism, or Neleman \& Tout's 2005
$\gamma$ parameter). And some of of the variety is due the use of
different assumed input parameters and distributions. But,
disturbingly, attempts by some teams (e.g. Mennekens et al. 2010) to reproduce
results of other teams by using the same recipes and inputs still show
significant discrepancies. Under this state of affairs, it appears
that the theoretical SD predictions for the DTD have not yet reached the 
point where they can be meaningfully compared to the observations. One generic
prediction that SD models do seem to make, however, is that the DTD
tends to drop off to zero well before a Hubble time. If this is
correct (and it may well not be; Phillip Podsiadlowski has often
pointed out that the BPS calculations also fail to predict adequate
numbers of other binary populations that we know do exist), 
this would mean that SD SNe Ia do
not play a role in producing the DTD tail clearly seen at long delays
in the observations. However, the present data cannot exclude also an SD 
contribution at short delays, present in tandem with a DD component that
produces the $t^{-1}$ power law DTD at long delays.

In summary, a host of measurements over the past few years have
revealed an increasingly clear picture of the SN Ia DTD. It is well
described by a power law of index $-1$, or somewhat steeper, going out
to a Hubble time. At delays of $< 1$~Gyr, this shape may continue,
or the slope may become somewhat shallower. The time-integrated SN Ia 
production efficiency is one SN Ia event for every $500 {\rm M_\odot}$
formed in stars, accurate to better than a factor of 2. The observed
DTD form is strikingly similar to the form generically expected, due
to fundamental gravitational wave physics, from Ron Webbink's (1984)
DD idea. The efficiency of SN Ia production by detailed models still
falls short of the observed number, by at least a factor of a few.
The competing SD model makes predictions that differ from the
observations both in DTD form and in the absolute numbers of
SNe. Given the disagreement among the SD calculations themselves, it is
not yet clear if this is a problem of the SD model or of its calculation.
But, keeping all these caveats in mind, the current picture appears
to be a ``thumbs up'' to Ron's model!



\begin{theacknowledgments}
I thank Ron Webbink and his co-birthday boys, 
 Ed van den Heuvel, and Peter Eggleton, for having
been such a source of inspiration to the field, and wish them many 
more years of invaluable contributions. I am grateful to the
organizers for a most enjoyable and illuminating meeting in such 
a pleasant and beautiful environment. The results presented here have
resulted from a decade of efforts to measure and understand SN rates
with my collaborators, students, and students-turned-collaborators: 
Carles Badenes, Alex Filippenko, Ryan Foley, 
Masataka Fukugita,
Avishay Gal-Yam, Or Graur, Raja Guhathakurta, Assaf Horesh, Buell Jannuzi,
Weidong Li, Filippo Mannucci, Eran Ofek, Dovi Poznanski, Paco Prada, 
Keren Sharon, Tomo
Totani,  and Naoki Yasuda; I thank them all.
This work is supported by a 
grant from the Israel Science Foundation. 
\end{theacknowledgments}



\bibliographystyle{aipproc}   


\IfFileExists{\jobname.bbl}{}
 {\typeout{}
  \typeout{******************************************}
  \typeout{** Please run "bibtex \jobname" to optain}
  \typeout{** the bibliography and then re-run LaTeX}
  \typeout{** twice to fix the references!}
  \typeout{******************************************}
  \typeout{}
 }

\end{document}